\documentclass{aa}  
\usepackage{pictex}
\usepackage{graphicx}
\usepackage{color}
\usepackage{natbib}
\usepackage{subfigure} 
\bibpunct{(}{)}{;}{a}{}{,}
\usepackage{txfonts}

\newcommand{\om}{\Omega_\mr m}
\newcommand{\sig}{\sigma_8}
\newcommand{\C}{\mathbf C}

\newcommand{\vxi}{\vec \xi}
\newcommand{\vxip}{\vec \xi_{\vpi}}
\newcommand{\vxipo}{\vec \xi_{\vpi_0}}

\newcommand{\vhxi}{\vec{\hat \xi}}

\newcommand{\vpi}{\vec \pi}

\newcommand{\mr}{\mathrm}

\newcommand{\tn}{\textnormal}
\newcommand{\be}{\begin{equation}}
\newcommand{\ee}{\end{equation}}

\newcommand{\nn}{\nonumber}
\newcommand{\beq}{\begin{eqnarray}}
\newcommand{\eeq}{\end{eqnarray}}
\newcommand{\ensav}[1]{\left\langle #1 \right\rangle}

\begin{document}

   \title{Dependence of cosmic shear covariances on cosmology}
\subtitle{Impact on parameter estimation}

   \author{T. Eifler \inst{1}, P. Schneider \inst{1} and J. Hartlap \inst{1}}
   \offprints{tim.eifler@astro.uni-bonn.de}

   \institute{1) Argelander-Institut f\"ur Astronomie, Universit\"at Bonn, Auf dem H\"ugel 71, D-53121 Bonn, Germany}

   \date{}

\abstract
{In cosmic shear likelihood analyses the covariance is most commonly assumed to be constant in parameter space. Therefore, when calculating the covariance matrix (analytically or from simulations), its underlying cosmology should not influence the likelihood contours.}
{We examine whether the aforementioned assumptions hold and quantify how strong cosmic shear covariances vary within a reasonable parameter range. Furthermore, we examine the impact on likelihood contours when assuming different cosmologies in the covariance. The final goal is to develop an improved likelihood analysis for parameter estimation with cosmic shear.}
{We calculate Gaussian covariances analytically for 2500 different cosmologies. In order to quantify the impact on the parameter constraints we perform a likelihood analysis for each covariance matrix and compare the likelihood contours. To improve on the assumption of a constant covariance, we use an adaptive covariance matrix, which is continuously updated according to the point in parameter space where the likelihood is evaluated. As a side-effect, this cosmology dependent covariance improves the parameter constraints. We examine this fact more closely using the Fisher-matrix formalism. In addition we quantify the impact of non-Gaussian covariances on the likelihood contours using a ray-tracing covariance derived from the Millennium simulation. In this ansatz we return to the approximation of a constant covariance matrix; in order to minimize the error due to this approximation, we develop the concept of an iterative likelihood analysis.}
{Covariances vary significantly within the considered parameter range. The cosmology assumed in the covariance has a non-negligible impact on the size of the likelihood contours. This impact increases with increasing survey size, increasing number density of source galaxies, decreasing ellipticity noise, and when taking non-Gaussianity into account. A proper treatment of this effect is therefore even more important for future surveys. In this paper we present methods to take cosmology dependent covariances into account.}
{}


\maketitle
%

\section{Introduction}
Cosmic shear, which was first detected in 2000 \citep{bre00,kwl00,wme00,wtk00}, has recently progressed to an important tool in cosmology. Latest results \citep[e.g.][]{wmh05,smw06,hmv06,ses06,het07,mrl07,fsh08} already indicate its great ability to constrain cosmological parameters which will be enhanced in the future by large upcoming surveys like Pan-STARRS, KIDS, DES, Euclid or LSST. The improved quality of cosmic shear data must be accompanied with an accurate data analysis, free of assumptions which bias the results. Obtaining appropriate covariances is a crucial issue in this context of a precision cosmology likelihood analysis. Several methods are suggested in the literature and have been applied to cosmic shear data. An analytic expression for covariances assuming a Gaussian shear field is derived in \cite{svk02} and confirmed in \cite{jse07} who use a power spectrum approach which significantly reduces the computational effort in the calculation. This analytic expression has been used for parameter estimation in many surveys \citep[e.g.][]{wmh05,smw06,hmv06}. However, the assumption of a Gaussian shear field breaks down on small scales; according to \cite{kil05} and \cite{svh07} non-linear effects already become important at angular scales $\lesssim$ 10 arcmin. To account for non-Gaussianity, \cite{svh07} invent a calibration factor which is derived from a comparison of Gaussian to ray-tracing covariances. An application of this method to real data can be found in \citep{fsh08}. A second approach is the derivation of the covariance matrix from the data \citep[e.g.][]{het07,mrl07}. Here, the covariance is calculated via field-to-field variation which involves a separation of the data set into many independent subsamples. This might lead to a loss of information on large scales if the survey is not sufficiently large. Third, one can estimate the covariance matrix from ray-tracing simulations, a method which circumvents the aforementioned loss in information. Although, in this method the covariance is again derived via field-to-field variation, we can choose a sufficiently large numerical simulation to create many independent subsamples of adequate size.\\
Note, that the last two methods involve an estimation process in the determination of the covariance matrix, which means that the inverse is biased and one has to correct for this effect \citep{and03,har07}. Nevertheless, deriving covariance matrices from ray-tracing simulations seems to be a promising method as it preserves all the information in the data and additionally takes the non-Gaussianity of the shear field into account.\\ 
The analytic expression and the ray-tracing covariance assume a specific cosmological model in their derivation. So far, cosmic shear likelihood analyses treat the covariance matrix as constant in parameter space, hence its underlying cosmology is assumed not to influence the parameter constraints. It is the intention of this paper to check for this assumption and in case it does not hold, to present an improved likelihood formalism for future surveys. \\
This paper is organized as follows. Section \ref{sec:basics} summarizes the basic theoretical background of the cosmic shear two-point correlation function (2PCF) and its corresponding covariance. In Sect. \ref{sec:varcov} we derive a scaling relation for covariances, which can be used for a fast calculation of covariances for arbitrary cosmology. Furthermore, we examine how strongly the covariance depends on its underlying cosmological model. The impact on parameter constraints when assuming a fixed cosmology in the covariance is subject of Sect. \ref{sec:varlike}, whereas we present improvements on this assumption in Sect. \ref{sec:varlike_adap}. Here, we consider a likelihood analysis with an adaptive covariance matrix, i.e. the covariance is calculated individually for each point in parameter space where the likelihood is evaluated. In addition, we outline the concept of an iterative covariance matrix, i.e. several likelihood analyses are performed, where the covariance is updated in every iteration according to the maximum likelihood parameter set. Here, we also examine the impact of non-Gaussian covariances on the likelihood contours using a ray-tracing covariance matrix derived from the Millennium Simulation. We conclude in Sect. \ref{sec:conc}.
 
\section{Data vectors and covariances of cosmic shear}
\label{sec:basics}
In this section we briefly review the basics of the cosmic shear two-point correlation function and its corresponding covariance matrix. For more details on this topic the reader is referred to \cite{bas01,svk02,svm02,kil04,jse07}. \\
To measure the shear signal we define $\vec \theta$ as the connecting vector of two points and specify tangential and cross-component of the shear $\gamma$ as
\be
\gamma_\mr t = - \mr{Re} \left( \gamma \mr e^{-2\mr i \varphi} \right) \qquad  \tn{and}  \qquad
\gamma_{\times} = - \mr{Im} \left( \gamma \mr e^{-2\mr i \varphi} \right) \;,
\ee 
where $\varphi$ is the polar angle of $\vec \theta$. The 2PCFs depend only on the absolute value of $\vec \theta$. They are defined in terms of the shear and can be related to the power spectra $P_{\mr E}$ and $P_{\mr B}$ \citep{svm02}
\beq
\label{eq:xifrome}
\xi_{\pm} (\theta) &\equiv& \langle \gamma_\mr t \gamma_\mr t  \rangle (\theta) \pm \langle \gamma_{\times} \gamma_{\times} \rangle (\theta) \\
\label{eq:xi+-}
&=&\int^{\infty}_0 \frac{\mr d\ell\;\ell}{2\pi} \, \mr J_{0/4}(\ell \theta)\, \,\left[P_\mr E(\ell) \pm P_\mr B(\ell)\right]\,,
\eeq
with $\mr J_n$ denoting the $n$-th order Bessel-function. In this paper we only consider E-modes, therefore we set  $P_\mr B=0$ from now on. Furthermore, we assume that the 2PCF is estimated in logarithmic bins $\vartheta$ of angular width $\Delta \vartheta$. The covariance of the 2PCF is defined as
\be
\label{eq:covxi}
\tn C_\xi \left( \vartheta_i, \vartheta_j \right) := \left \langle
\left( \xi_{\pm} (\vartheta_i)\, - \,\hat \xi_{\pm} (\vartheta_i)\right) \left( \xi_{\pm}(\vartheta_j)\,-\, \hat \xi_{\pm}(\vartheta_j)\right) \right \rangle.
\ee
We neglect the index $\xi$ in the covariance for the rest of the paper as we only consider covariances of the 2PCF. As one already sees from (\ref{eq:covxi}) the 2PCF has four different covariances, denoted as $\tn C_{++}$, $\tn C_{+-}$, $\tn C_{-+}$, $\tn C_{--}$. Only three of them are independent since $\tn C_{+-}(\vartheta_i ,\vartheta_j)= \tn C_{-+}(\vartheta_j,\vartheta_i)$. Assuming a Gaussian shear field the covariance of the 2PCF can be calculated analytically \citep{svk02,jse07}. There, the covariance is decomposed into three terms, namely the cosmic variance term (V), the pure shot noise term (S), and the mixed term (M)
\beq
\label{eq:cov++}
\tn C_{++}(\vartheta_i ,\vartheta_j) &=& V_{++} + M_{++}+ S \,, \\
\label{eq:cov--}
\tn C_{--}(\vartheta_i ,\vartheta_j) &=& V_{--} + M_{--}+ S \,,\\
\label{eq:cov+-}
\tn C_{+-}(\vartheta_i ,\vartheta_j) &=& V_{+-} + M_{+-} \,.
\eeq
The pure shot noise term vanishes in case of $\tn C_{+-}$ and only contributes to the diagonal of $\tn C_{++}$ and $\tn C_{--}$. It can be calculated as
\be
\label{eq:covS}
S = \frac{\sigma^4_\epsilon}{2 \pi \vartheta_i  \Delta \vartheta_i  A  {\bar n}^2} \, \delta_{\vartheta_i \vartheta_j} \, ,
\ee
where $A$ denotes the solid angle of the data field, $\sigma_\epsilon$ is the intrinsic ellipticity dispersion, and $\bar n$ the number density of source galaxies. The cosmic variance term (V) and the mixed term (M) can be either calculated using the power spectrum or the 2PCF. According to \cite{jse07} the power spectrum approach leads to the following expressions 
\beq
\label{eq:covV}
V_{\pm \pm}&=& \frac{1}{\pi A} \int_0^{\infty} \tn d \ell \, \ell \, \mr J_{0/4} (\ell \vartheta_i) \, \mr J_{0/4}(\ell \vartheta_j) \, P^2_\mr E (\ell) \,, \\ 
\label{eq:covM}
M_{\pm \pm}&=& \frac{\sigma^2_\epsilon}{\pi A \bar n} \int_0^{\infty} \tn d \ell \, \ell \, \mr J_{0/4} (\ell \vartheta_i) \, \mr J_{0/4}(\ell \vartheta_j) \,P_\mr E (\ell) \,.
\eeq
The corresponding expressions for $V$ and $M$ using the 2PCF are derived in \cite{svk02}. In this paper we only need the expressions for the mixed term, which read  
\beq
\label{eq:covMreal1}
M_{++}&=& \frac{2 \sigma^2_\epsilon}{\pi A n} \int_0^\pi \mr d \varphi \, \xi_+ (|\vec \phi|)
\,, \\
\label{eq:covMreal2}
M_{--}&=& \frac{2 \sigma^2_\epsilon}{\pi A n} \int_0^\pi \mr d \varphi \, \xi_+ (|\vec \phi|) \, \cos (4 \varphi) \,,\\
\label{eq:covMreal3}
M_{+-}&=& \frac{2 \sigma^2_\epsilon}{\pi A n} \int_0^\pi \mr d \varphi \left[ \sum^4_{k=0} \left( 4 \atop k \right) (-1)^k \vartheta^k_i \, \vartheta^{4-k}_j \cos(k \varphi) \right] \nn \\
&\quad& \qquad \times \, (|\vec \phi|)^{-4} \, \xi_- (|\phi|) \, \cos (4 \varphi) \, ,
\eeq
where we denote $|\vec \phi|= \sqrt{ \vartheta^2_i+\vartheta^2_j- 2 \vartheta_i \vartheta_j \cos \varphi}$.
\section{Variation of covariances in parameter space}
\label{sec:varcov}
We select a two-dimensional parameter grid with 50 $\times$ 50 gridpoints of $\om \in [0.2;0.4]$ and $\sig \in [0.6;1.0]$. For each grid point we calculate a covariance analytically using (\ref{eq:cov++}) - (\ref{eq:covM}). The shear power spectra $P_\mr E$ are obtained from the density power spectra $P_\delta$ employing Limber's equation. To derive $P_\delta$ we assume an initial Harrison-Zeldovich power spectrum ($P_{\delta}(k) \propto k^{n_\mr s}$ where $n_\mr s=1$) with the transfer function from \cite{ebw92}. For the calculation of the non-linear evolution we use the fitting formula of \cite{sm03}. Throughout this paper, we assume a flat universe and fix all cosmological parameters except $\om$ and $\sig$, more precisely $H_0=0.73$ and $\Omega_\mr b=0.04$. These values for $H_0$ and $\Omega_\mr b$ together with $\om=0.25$ and $\sig=0.9$ define our fiducial cosmological model, which we have chosen similar to the cosmology of the Millennium Simulation \citep{swj05} for a later comparison of Gaussian and ray-tracing covariances. We assume all source galaxies to be at redshift $z_0=1.0$. Using a redshift distribution instead would not change our results qualitatively. In addition to cosmology, the covariance depends on survey parameters. The scaling relations given in Sect. \ref{sec:varcov} are generally valid and independent of survey parameters. In case of the likelihood analyses in Sects. \ref{sec:varlike} and \ref{sec:varlike_adap} we choose, unless stated otherwise, an intrinsic ellipticity noise of $\sigma_\epsilon=0.4$, a number density of source galaxies of $\bar n=10/\tn{arcmin}^2$ (similar to the values of the Dark Energy survey), and a survey which covers $A=900$ deg$^2$. The angular scale of the 2PCF data vector for which we calculate the covariances covers a range from 0.1 arcmin to 180 arcmin, which is divided into 50 logarithmic bins.

\subsection{A fast method to calculate covariances for arbitrary $\om$ and $\sig$}
\label{sec:varcov_calc}
\begin{figure}
\resizebox{\hsize}{!}{\includegraphics[width=8cm]{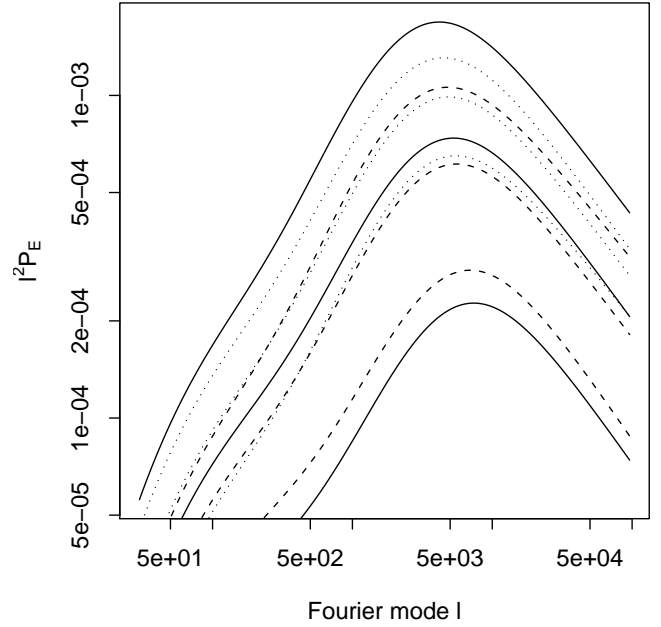}}
\caption{The dimensionless shear power spectrum $ \ell^2 P_\mr E$. The solid curves correspond to variation in $\om$ and $\sig$: $\om=0.2$, $\sig=0.6$ (\textit{lower}), $\om=0.3$, $\sig=0.8$ (\textit{middle}), $\om=0.4$, $\sig=1.0$ (\textit{top}). The dashed curves show variation in $\sig$ with $\om=0.25$: $\sig=0.6$ (\textit{lower}), $\sig=0.8$ (\textit{middle}), $\sig=1.0$ (\textit{top}). The dotted curves show variation in $\om$ with a constant $\sig=0.9$: $\om=0.2$ (\textit{lower}), $\om=0.3$ (\textit{middle}), $\om=0.4$ (\textit{top}).}
\label{fi:power}
\end{figure}
From (\ref{eq:covV}) and (\ref{eq:covM}) one directly sees that the covariance matrix depends on the cosmological model, which enters with the power spectrum $P_\mr E$. Figure \ref{fi:power} illustrates the change in $P_\mr E$ when varying only $\om$, or $\sig$, and both parameters simultaneously; we see that it increases with $\om$ as well as with $\sig$.\\
For a given cosmological model we can calculate the covariance directly from (\ref{eq:cov++}) - (\ref{eq:covM}). Performing this calculation for many sets of parameters is time-consuming; hence we seek a scaling relation, which relates the covariances of an arbitrary cosmology $\C_{\vpi}$ to a reference model $\C_{\vpi_0}$. A basic theorem in statistics states \citep[e.g.][]{and03}, that if there is a relation between two data vectors $\vec x$ and $\vec y$ which reads $\vec y= \mathbf A \vec x$ ($\mathbf A$ being a  matrix), the relation of the covariances of $\vec x$ and $\vec y$ can be written as
\beq
\label{eq:covscaletheo}
\C_{\vec y} &=& \ensav{(\vec y-\ensav{\vec y})(\vec y-\ensav{\vec y})^\mr t} \nn \\
	  &=& \ensav{(\mathbf A \vec x -\ensav{\mathbf A \vec x})(\mathbf A \vec x-\ensav{\mathbf A \vec x})^\mr t} \nn \\
&=&\mathbf A \C_{\vec x} \mathbf A^t \,.
\eeq
In this derivation $\mathbf A$ must be independent of the ensemble average. If we apply the above ansatz to the 2PCF, it seems reasonable to define a scaling relation for parameter dependent covariances as
\be
\label{eq:covscale2PCF}
\C_{\vxip} = \mathbf A \C_{\vxipo} \mathbf A^{\mr t} \,,
\ee
where we can calculate the scaling matrices $ \mathbf A$ using the 2PCF
\be 
diag(\mathbf A) = \vxip/\vxipo \,.
\ee 
In contrast to a covariance matrix, the 2PCF can be calculated extremely fast for many different cosmologies via (\ref{eq:xi+-}). Hence, it would be a fast and convenient method to calculate the covariance for a reference cosmology and then apply (\ref{eq:covscale2PCF}) to obtain covariances for arbitrary cosmological parameters. Unfortunately, we cannot transfer this method directly to the cosmic shear case. Recall, that the 2PCF is derived from the measured ellipticities of galaxies. \cite{svm02} have shown that the intrinsic ellipticity terms cancel out in the derivation of the 2PCF estimator, hence the 2PCF is defined only in terms of the shear. In contrast, the 2PCF covariance does not only consist of terms coming from the shear, but has additional noise terms which arise from the intrinsic ellipticity of galaxies. The pure shot noise term (\ref{eq:covS}) is independent of cosmology and, as can be seen from (\ref{eq:covMreal1}-\ref{eq:covMreal3}), the mixed term cannot be scaled with relation (\ref{eq:covscale2PCF}), which is quadratic in the 2PCF.\\
However, in the limit of a noise-free covariance, i.e. considering only the cosmic variance term, a scaling relation similar to (\ref{eq:covscale2PCF}) exists. We explicitly prove this below, in particular, we show that the scaling matrices are independent of the ensemble average. The cosmic variance term can be calculated via (\ref{eq:covV}). Cosmology only enters with the power spectrum, hence the relation of $\vpi$ to $\vpi_0$ can be described as $P_\mr E (\ell, \vpi ) = a (\ell, \vpi)  P_\mr E (\ell, \vpi_0 )$. Using this relation we transform the cosmic variance term (\ref{eq:covV}) for given bins $\vartheta_i, \vartheta_j$ as follows
\beq
\label{eq:Vscale1}
V_{\pm \pm}(\vpi)&=&  \frac{1}{\pi A} \int_0^{\infty} \tn d \ell \, \ell \, \mr J_{0/4} (\ell \vartheta_i) \, \mr J_{0/4}(\ell \vartheta_j) \, P^2_\mr E (\ell, \vpi ) \nn \\
&=& \frac{1}{\pi A} \sum_{\bar{\ell}} \Delta \bar{\ell} \, \bar{\ell} \, \mr J_{0/4} (\bar{\ell} \vartheta_i) \, \mr J_{0/4}(\bar{\ell} \vartheta_j) \,a^2( \bar{\ell}, \vpi) P^2_\mr E ( \bar{\ell}, \vpi_0) , 
\eeq
where we discretize the integral into a sum of $\bar{\ell}$-bins. Next we insert equation (26) of \cite{jse07} \citep[see also][]{kai98} but with $\sigma_\epsilon=0$
\be
\label{eq:ps_ensemble}
\ensav{\Delta P_\mr E (\bar{\ell}) \Delta P_\mr E (\bar{\ell}')} =\frac{4 \pi}{A \bar{\ell} \Delta \bar{\ell}} \, P^2_\mr E (\bar{\ell})\, \delta_{\bar{\ell} \bar{\ell}'} \, ,
\ee
to rewrite (\ref{eq:Vscale1}) as
\beq
\label{eq:Vscale2}
V_{\pm \pm}(\vpi) &=& \frac{1}{4 \pi^2} 
\sum_{\bar{\ell}, \bar{\ell'}} \Delta \bar{\ell}^2  \bar{\ell}' \bar{\ell} \,\mr J_{0/4} (\bar{\ell} \vartheta_i)\, \mr J_{0/4}(\bar{\ell}'  \vartheta_j)\, a(\bar{\ell}, \vpi) \, a(\bar{\ell}', \vpi)\nn \\ &\quad& \times  \ensav{\Delta P_\mr E (\bar{\ell}, \vpi_0) \Delta P_\mr E (\bar{\ell}',\vpi_0)}\,. 
\eeq
The mean value theorem guarantees that there exist values $\bar{a}(\vartheta_i, \vpi)$, $\bar{a}(\vartheta_j, \vpi)$ such that (\ref{eq:Vscale2}) becomes
\beq
V_{\pm \pm}(\vpi) &=& \frac{\bar{a}(\vartheta_i, \vpi) \, \bar{a}(\vartheta_j, \vpi)}{4 \pi^2} 
 \ensav{\int_0^{\infty} \mr d \ell \, \ell \, \mr J_{0/4} (\ell \vartheta_i) \, \Delta P_\mr E (\ell,\vpi_0) \right.  \nn \\ 
&\quad& \times \left. \int_0^{\infty} \mr d \ell'  \ell' \, \mr J_{0/4} (\ell' \vartheta_j) \,\Delta P_\mr E (\ell',\vpi_0)}    \nn \\
&=& \bar{a}(\vartheta_i, \vpi)\, \bar{a}(\vartheta_j, \vpi)\, V_{\pm \pm}(\vpi_0)\, ,
\eeq
where we consider the limit $\Delta \bar{\ell} \longrightarrow 0$ in the first step. Comparing the expressions of $V_{\pm \pm}(\vpi)$ and $V_{\pm \pm}(\vpi_0)$ we can calculate the scaling factors as 
\be
\label{eq:scaling_factor}
\bar{a}_{\pm \pm} (\vartheta_i, \vpi) = \frac{\int_0^{\infty} \mr d \ell \, \ell \, \mr J_{0/4} (\ell  \vartheta_i)  P_\mr E (\ell,\vpi)}{\int_0^{\infty} \mr d \ell \, \ell \, \mr J_{0/4} (\ell  \vartheta_i)  P_\mr E (\ell,\vpi_0)} =\frac{\xi(\vartheta_i, \vpi)}{\xi(\vartheta_i, \vpi_0)}\,,
\ee
where we inserted (\ref{eq:xi+-}) in the last step. This provides a fast and convenient method to scale the cosmic variance term in parameter space, due to the fact we can use a computationally efficient Hankel transformation for the calculation of the 2PCF.\\
From (\ref{eq:covMreal1}) - (\ref{eq:covMreal3}) we see that the mixed term $M_{\pm \pm}$ scales linearly with the 2PCF which prevents a scaling relation similar to (\ref{eq:covscale2PCF}). Fortunately, the direct calculation of the linear term via (\ref{eq:covM}) is comparatively fast, therefore, the scaling relation for the cosmic variance term already reduces the computational costs significantly.\\
Nonetheless, we numerically derive a fit-formula for the linear term based on the following expression
\be
\label{eq:fit}
M_{\pm \pm}(\vartheta_i,\vartheta_j,\vpi) =  M_{\pm \pm} (\vartheta_i, \vartheta_j,\vpi_0) \left( \frac{\om}{0.25}\right)^{\alpha} \left( \frac{\sig}{0.9} \right)^{\beta} \,.
\ee
The structure of this fit-formula is motivated by the intention to use as few fit-parameters as possible; additionally we require that in the limit of the fiducial model, $M_{\pm \pm}(\vartheta_i,\vartheta_j,\vpi)= M_{\pm \pm} (\vartheta_i, \vartheta_j,\vpi_0)$ must hold. The fit-parameters $\alpha$ and $\beta$ vary depending on the scale $\vartheta_i,\vartheta_j$ and are different for the different parts of the covariance matrix, $\C_{++}$, $\C_{--}$, and $\C_{+-}$. The tables with $\alpha$ and $\beta$ are available on the internet\footnote{http://www.astro.uni-bonn.de/$\sim$teifler/fit-parameters.pdf}. 
\subsection{Variation of the inverse covariance with $\om$ and $\sig$}
\label{sec:varcov_para}
From the variation of the power spectrum with $\om$ and $\sig$ (Sect. \ref{sec:varcov_calc}) it is clear that covariances vary with respect to comological parameters. For simplicity and in order to increase the readability of the following sections we refer to this variation as CDC-effect (CDC $\equiv$ Cosmology Dependent Covariances). 
In order to examine the CDC-effect more closely, recall that the structure of the covariance is given by
\be
\C=\left( \begin{array}{cccccccc}
		\multicolumn{7}{c|}{\rule[-3mm]{0mm}{8mm}\textbf{$\C_{++}$}}& \multicolumn{1}{c}{\textbf{$\C_{+-}$}}\\
		\hline
		\multicolumn{7}{c|}{\rule[1mm]{0mm}{3mm}\textbf{$\C_{+-}^\mr t$}}& \multicolumn{1}{c}{\textbf{$\C_{--}$}}
		 \end{array} \right) \nn
\ee
and the individual parts are calculated from (\ref{eq:cov++}) - (\ref{eq:covM}). From these equations we see that the covariances are filtered versions of the power spectrum, either filtered by a product of $\mr J_0$'s (in case of $\C_{++}$), $\mr J_4$'s ($\C_{--}$), or a combination of both ($\C_{+-}$). The strength of the CDC-effect depends on these filter functions, as they determine which parts of the power spectrum are sampled. A change in $\om$ and $\sig$ affects all scales of the power spectrum almost similarly (see Fig. \ref{fi:power}) therefore, the CDC-effect for the individual parts of $\C$ is also similar. However, this might change when considering different cosmological parameters, e.g. the shape parameter $\Gamma$. A change in $\Gamma$ rotates the power spectrum. The covariances are integrals over $P_\mr E$, and depending on the filter function, the change in $P_\mr E$ can average out. A second argument why the individual covariance parts have different sensitivity to the CDC-effect is that $\C_{+-}$ is not affected from shot noise, hence a change in cosmology has a stronger impact on $\C_{+-}$ compared to $\C_{++}$ and $\C_{--}$.\\
In order to quantify the CDC-effect we examine the trace of the inverse covariance matrix $\C^{-1}$. The trace of the covariance itself is an improper measure for this effect, as it depends on the binning, which can be seen from (\ref{eq:covS}). The trace of $\C$ becomes arbitrarily large when decreasing the bin width. In contrast, we checked numerically that for the trace of $\C^{-1}$ binning effects are negligible, once one has exceeded a minimum bin number. More precisely, once the bin width of the 2PCF data vector is small enough that discretization effects are unimportant, the trace of $\C^{-1}$ hardly changes for different binning. \\
\begin{figure}
\resizebox{\hsize}{!}{\includegraphics[width=8cm]{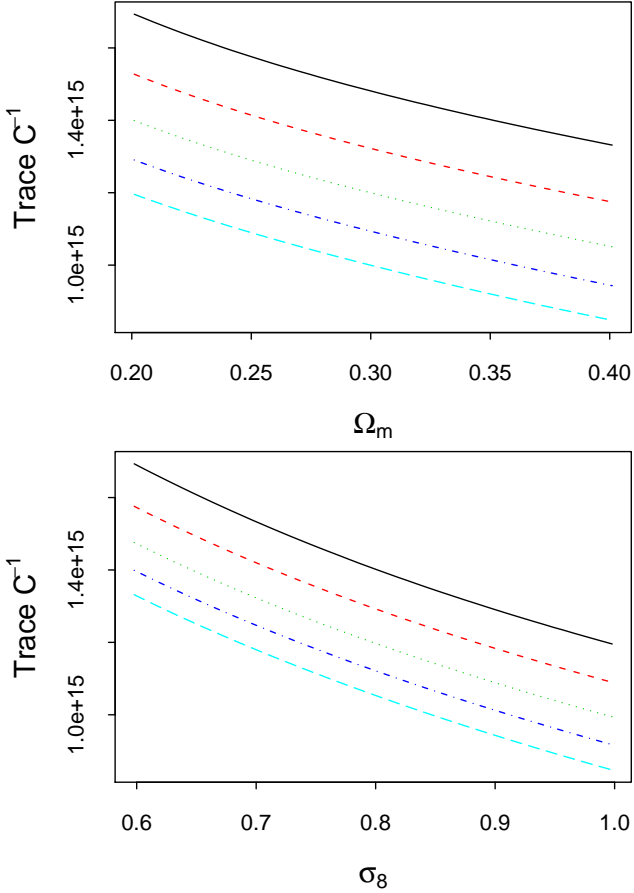}}
\caption{The trace of the inverse covariance matrix $\C^{-1}$ depending on $\om$ (\textit{top}), the individual lines in each figure correspond to (from top to bottom) $\sig = [0.6, 0.7, 0.8, 0.9, 1.0]$. The \textit{lower} panel shows the dependence on $\sig$, the individual lines corresponding to (from top to bottom) $\om = [0.2, 0.25, 0.3, 0.35, 0.4]$.}
\label{fig:trace_tot}
\end{figure}
Figure \ref{fig:trace_tot} shows the trace of the inverse covariance matrix depending on $\om$ for various constant $\sig$ (top) and vice versa (bottom). Here, we normalize the survey size to $A= 1\, \mr{deg}^2$; the other survey parameters are $\sigma_\epsilon=0.4$ and $\bar n=10/\tn{arcmin}^2$. We postpone a detailed analysis of how survey parameters influence the CDC-effect to Sect. \ref{sec:varlike}. Qualitatively the result does not change for different survey parameters; the trace of $\C^{-1}$ decreases with increasing $\om$ or $\sig$. \\
In addition, we perform a singular value decomposition (SVD) for each inverse covariance matrix. For the case of a symmetric and positive definite matrix, such as the inverse covariance matrix, an SVD yields the eigenvalues in decreasing order. For arbitrary $i$, we find that the $i$-th eigenvalue decreases when increasing $\om$ or $\sig$. The strength of the CDC-effect, i.e. the gradient of the traces, depends on the considered point in parameter space. 

\section{Impact of the CDC-effect on parameter estimation}
\label{sec:varlike}
\subsection{Basics of the likelihood analysis}
\label{sec:varlike_basics}
Throughout the whole likelihood analysis we assume the $\Lambda \mr{CDM}$ model. We define the posterior likelihood $p(\vpi|\vxi)$ for the case of a 2PCF data vector as  
\be
\label{eq:postlike}
p(\vpi|\vxi)=\frac{p(\vxi|\vpi) }{p(\vxi)} \, p(\vpi) \, ,
\ee
where $p(\vpi)$ denotes the prior probability density, $p(\vxi|\vpi)$ is the likelihood and $p(\vxi)$ denotes the evidence. The prior usually contains knowledge on the parameter vector $\vpi$ coming from former experiments. Here, we assume flat priors with cutoffs, which means $p(\vpi)$ is constant for all parameters inside a fixed interval (i.e. $\om \in [0.2;0.4]$, $\sig \in [0.6;1.0]$) and $p(\vpi)=0$ else. We assume that $\vxi$ is normally distributed in parameter space, hence our likelihood $p(\vxi|\vpi)$ can be written as 
\be
\label{eq:likefunc}
p(\vxi |\vpi) = \frac{\exp \, \left[ -\frac{1}{2} \, \left( (\vxi_{\vpi} - \vhxi)^\mr t \;\C^{-1} \;( \vxi_{\vpi} - \vhxi)\right) \right]}{(2 \pi)^{d/2} \; |\C|^{\frac{1}{2}}}  \, ,
\ee
where $\vhxi$ denotes the mean data vector, $\vxi_{\vpi}$ the model data vector, $d$ is the dimension of the data vectors, hence $|\C|$ is the determinant of a $d \times d$ covariance matrix. Note that the 2PCF data vector consists of two parts ($\vec \xi_+$ and $\vec \xi_-$),  each with $d/2$ bins. The evidence is a normalization obtained by integrating the likelihood over the considered parameter space
\be
\label{eq:evidence}
p(\vxi)= \int \mr d \vpi \, \frac{\exp \left[- \frac{1}{2}\, \left((\vxi_{\vpi} - \vhxi)^\mr t \;\C^{-1} \;( \vxi_{\vpi} - \vhxi) \right) \right]}{(2 \pi)^{d/2} \; |\C|^{\frac{1}{2}}} \,.
\ee
In our case we calculate $\vhxi$ from $P_\mr E$ via (\ref{eq:xi+-}) assuming our fiducial cosmology; $\vxi_{\vpi}$ is calculated similarly but its cosmological model varies according to the considered point in parameter space. The result of a likelihood analysis is usually summarized in contour plots. In a Bayesian approach, these likelihood contours represent so-called credible regions, i.e. a region in parameter space, where the true parameter is located with a probability of 68\%, 95\%, 99,9\%, respectively. In addition, we quantify the size of these credible regions through the determinant of the second-order moment of the posterior likelihood \citep[see][]{kil04} 
\be
\label{eq:quadrups}
\mathcal Q_{ij} \equiv \int \tn d^2 \vpi p(\vpi|\vxi) \; (\pi_i-\pi_i^{\mr f})(\pi_j -\pi_j^{\mr f}),
\ee
with $\pi_1$ and $\pi_2$ as the varied parameters, $\pi_i^{\mr f}$ as the parameter of the fiducial model. The determinant is given by
\be
\label{eq:detquadrups}
q= \sqrt{|\mathcal Q_{ij}|} = \sqrt{\mathcal Q_{11} \mathcal Q_{22} - \mathcal Q_{12}^2}.
\ee 
Smaller credible regions in parameter space correspond to a smaller value of $q$. In this paper all $q$ are given in units of $10^{-4}$.
\begin{figure*}
  \includegraphics[width=18cm]{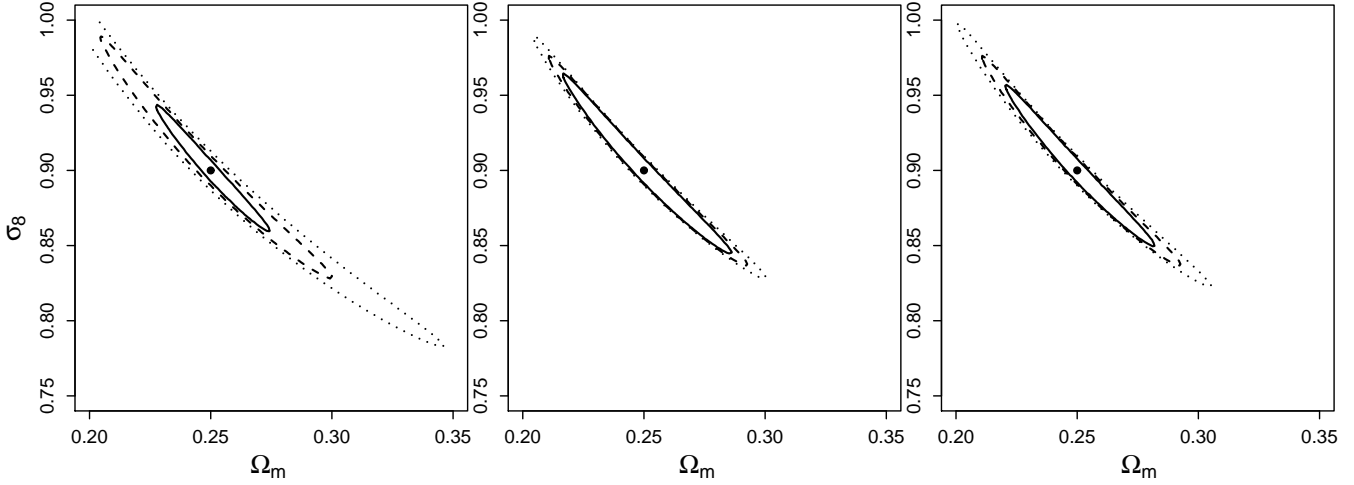}
        \caption{The 95\%-credible intervals obtained from likelihood analyses with different cosmological models assumed in their covariance matrix. The left panel corresponds to the following covariance parameters: $\Omega_\mr m=0.2$, $\sig=0.6$ (\textit{solid}), $\Omega_\mr m=0.25$, $\sig=0.9$ (\textit{dashed}), and $\Omega_\mr m=0.4$, $\sig=1.0$ (\textit{dotted}). The middle panel shows the deviation which occurs when restricting the range of possible covariance models to the 68\% confidence interval of the WMAP5 analysis, i.e.  $\Omega_\mr m=0.237$, $\sig=0.74$ (\textit{solid}), $\Omega_\mr m=0.259$, $\sig=0.796$ (\textit{dashed}), and $\Omega_\mr m=0.274$, $\sig=0.85$ (\textit{dotted}). The right panel shows the same analysis but for the 95\% confidence interval of the WMAP5 analysis, i.e. $\Omega_\mr m=0.226$, $\sig=0.70$ (\textit{solid}), $\Omega_\mr m=0.237$, $\sig=0.74$ (\textit{dashed}), and $\Omega_\mr m=0.288$, $\sig=0.885$ (\textit{dotted}).}
         \label{fig:like_constant}
   \end{figure*}
\subsection{Results of the likelihood analysis}
\label{sec:varlike_para}
\begin{figure}
\resizebox{\hsize}{!}{\includegraphics[width=8cm]{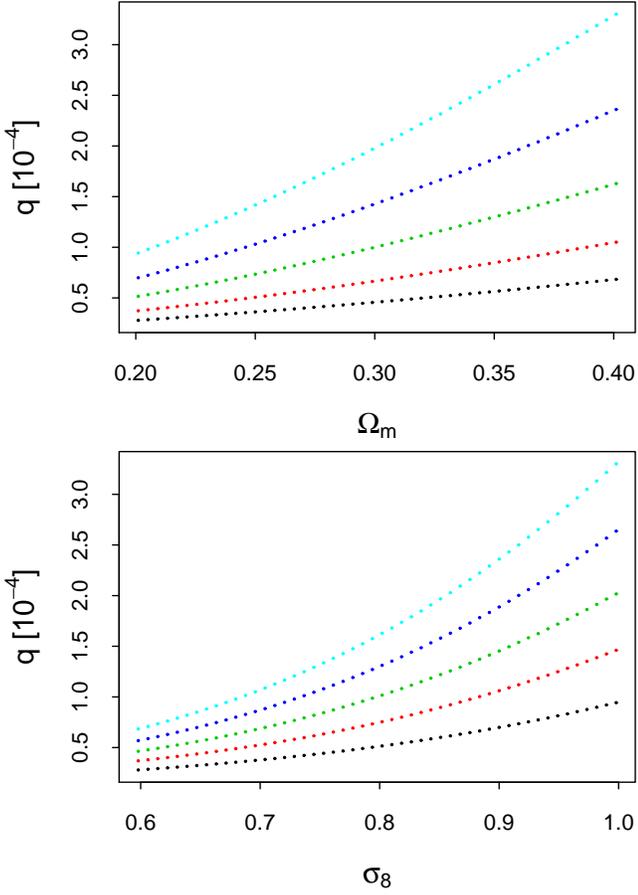}}
\caption{The values of $q$ depending on $\om$ (\textit{top}), the individual lines in each figure correspond to (from top to bottom) $\sig = [0.6, 0.7, 0.8, 0.9, 1.0]$. The \textit{lower} panel shows the dependence on $\sig$, the individual lines corresponding to (from top to bottom) $\om = [0.2, 0.25, 0.3, 0.35, 0.4]$.}
\label{fig:q_2500}
\end{figure}

In Sect. \ref{sec:varcov} we calculate 2500 covariances covering a parameter range of $\om \in [0.2;0.4]$ and $\sig \in [0.6;1.0]$. Here, we want to examine how the CDC-effect influences the likelihood contours, hence for each of the 2500 covariance matrices we perform a likelihood analysis. In these analyses, we consider the same parameter space, similar priors, similar $\vhxi$ and $\vxi_{\vpi}$, only the covariance in (\ref{eq:likefunc}) is changed. The left panel of Fig. \ref{fig:like_constant} shows the 95\%-credible intervals when choosing $\om=0.2$ and $\sig=0.6$ (solid), $\om=0.4$, and $\sig=1.0$ (dotted) as a model for the covariance matrix. We compare these to the (dashed) case when the covariance is calculated from the fiducial model ($\om=0.25$, $\sig=0.9$). These examples illustrate that assuming different cosmologies in the covariance can significantly broaden or narrow the likelihood contours. As expected from the foregoing analysis of the inverse covariance traces (Sect. \ref{sec:varcov}) the contours broaden for increasing $\om$ and $\sig$.\\
Without any information which cosmology to choose in our covariance matrix, it is reasonable to include prior information coming from other cosmological probes into our covariance cosmology. The middle panel of Fig. \ref{fig:like_constant} shows the 95\% credible intervals when calculating the covariance from the minimum, mean, and maximum values of the 68\% confidence region of the recent WMAP 5-years analysis \citep{kdn08}. Compared to the left panel the deviation of the contours reduces significantly, nevertheless it is still noticeable and cannot be neglected in a precision cosmology analysis. Similarly, the right panel shows the impact of the CDC-effect when calculating the covariance from parameters within the  95\% confidence region of the recent WMAP5 analysis. For a better comparison we calculate the values of $q$ (Sect. \ref{sec:varlike_basics}) for all contour plots and summarize them in Table \ref{tab:qvalues}. Restricting the possible cosmologies for the covariance to the 68\% contour region of the WMAP5 analysis, the values of $q$ deviate by a factor of $\approx  1.84$. This factor increases to $\approx 2.76 $ when considering the minimum and maximum values of the 95\% confidence region of the WMAP5 constraints. In Fig. \ref{fig:q_2500} we show the values of $q$ for all 2500 likelihood analyses depending on $\om$ (top) and $\sig$ (bottom). Similar to the parameter dependence of the inverse covariances in Sect. \ref{sec:varcov}, the strength of the CDC-effect, i.e. the gradient of the 
curves in Fig. \ref{fig:q_2500}, depends on the considered point in parameter space.  At the fiducial model we calculate $(\partial q/\partial \om)_\mr{fid} = 7.5$, whereas in case of $\sig$ we find $(\partial q/\partial \sig)_\mr{fid} = 3.5$.

\begin{table}
\caption{Values of $q$ for different covariance models}
\centering
\label{tab:qvalues}
\begin{tabular}{l l}\hline \hline
parameters used for the covariance& $q$ $[10^{-4}]$ \\ \hline
$\om = 0.25$, $\sig=0.9$ & 1.03\\ 
$\om = 0.2$, $\sig=0.6$& 0.28 \\
$\om = 0.4$, $\sig=1.0$& 3.30\\ \hline
$\om =  0.259$, $\sig=0.796$ (WMAP5 68 \% CL mean)&0.75\\ 
$\om =  0.237$, $\sig=0.740$ (WMAP5 68 \% CL min)& 0.56\\ 
$\om =  0.274$, $\sig=0.850$ (WMAP5 68 \% CL max)& 1.02\\ \hline
$\om =  0.226$, $\sig=0.700$ (WMAP5 95 \% CL min)& 0.45\\ 
$\om =  0.288$, $\sig=0.885$ (WMAP5 95 \% CL max)& 1.24\\ \hline
\end{tabular}
\end{table}

\subsection{Impact of survey parameters on the CDC-effect}
\label{sec:varlike_survey}
\begin{figure}
\resizebox{\hsize}{!}{\includegraphics[width=8cm]{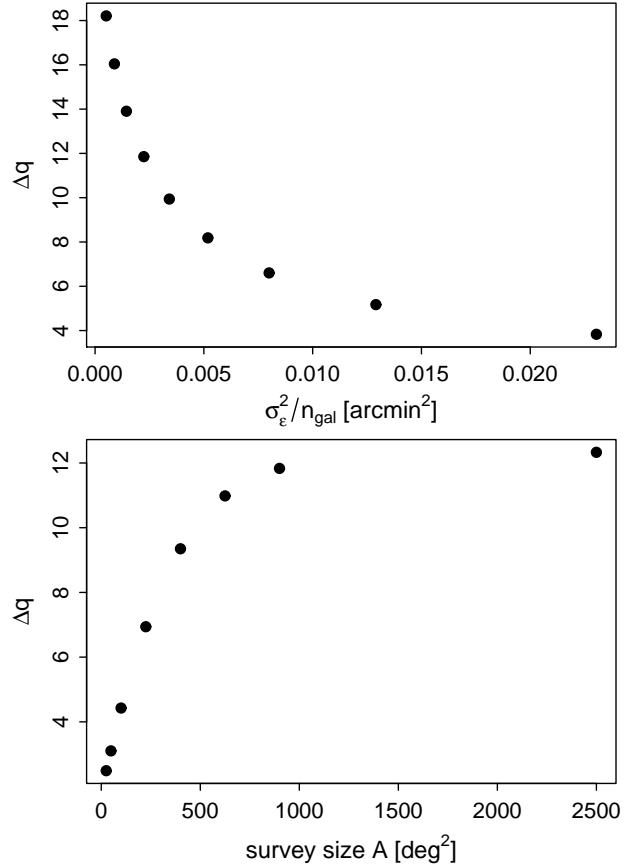}}
\caption{The ratio of maximum to minimum value of $q$ depending on the ratio $\sigma_\epsilon^2/\bar n$ (\textit{upper panel}) and depending on the survey size $A$ (\textit{lower panel}).}
\label{fi:survey1}
\end{figure}
In the last section we have shown, that the CDC-effect non-negligibly affects the likelihood contours. However, we only quantify this for one specific set of survey parameters. In this section we examine how the impact of the CDC-effect on likelihood contours depends on survey parameters, namely survey size $A$, ellipticity dispersion $\sigma_\epsilon$, and number density of source galaxies $\bar n$, where in case of the latter two only the combination $\sigma_\epsilon^2/\bar n$ is of interest. We perform likelihood analyses for 9 different combinations of $\sigma_\epsilon^2/\bar n$ and 8 different survey sizes. The strength of the CDC-effect is quantified by the ratio of maximum to minimum value of $q$, which occur within the considered range of $\om$ and $\sig$, we define $\Delta q=q(\vpi_\mr{max})/q(\vpi_\mr{min})$. The minimum $q$ is obtained when choosing the minimum parameter set in the calculation of the covariance, i.e. $\vpi_\mr{min}=(\om=0.2, \sig=0.6)$. Correspondingly, choosing the maximum parameter set $\vpi_\mr{max}=(\om=0.4, \sig=1.0)$ results in the maximal $q$. The values of $q$ represent the size of credible intervals, hence $\Delta q$ can be interpreted as their ratio. \\
Unfortunately, it is not possible to derive an analytical expression for the relation between $\Delta q$ and the survey parameters. From (\ref{eq:covS}) - (\ref{eq:covM}) we see that the individual covariance terms scale differently with $\sigma_\epsilon^2/\bar n$. This already prohibits an analytically derived relation between $\Delta q$ and $\sigma_\epsilon^2/\bar n$. Considering the survey size $A$, (\ref{eq:covS}) - (\ref{eq:covM}) imply that the total covariance scales with $1/A$. When comparing two (inverse) covariances with different cosmologies by taking their ratio, the survey size cancels, suggesting the strength of CDC-effect to be independent of $A$. However, when considering the likelihood, the inverse covariance enters in the exponent, furthermore the values of $q$ are an integral over the posterior likelihood. This non-linearity in the inverse covariance causes that the strength of the CDC-effect varies with the survey size. An analytic expression of this dependence cannot be derived, for similar reasons as for  the case of $\sigma_\epsilon^2/\bar n$. We therefore calculate $\Delta q$ depending on the survey parameters numerically.\\
The upper panel of Fig. \ref{fi:survey1} shows $\Delta q=q(\vpi_\mr{max})/q(\vpi_\mr{min})$ as a function of $\sigma_\epsilon^2/\bar n$. The ratio $\Delta q$ changes from 4 to 18 over the considered interval of $\sigma_\epsilon^2/\bar n$. When increasing the survey size $A$ (Fig. \ref{fi:survey1} , lower panel), we find that the impact of the CDC-effect increases from $\Delta q=2.5$ (for a 25 deg$^2$ survey) up to $\Delta q=12.3$ (for a 2500 deg$^2$ survey). Note that the size of the likelihood contours, hence the values of $q$ themselves, decrease with decreasing $\sigma_\epsilon^2/\bar n$ and increasing $A$. In contrast, $\Delta q$ increases with decreasing $\sigma_\epsilon^2/\bar n$ and increasing $A$. Hence relatively, the CDC-effect becomes more important when increasing the survey size or when decreasing the ratio $\sigma_\epsilon^2/\bar n$.

\section{Likelihood analysis with a model dependent covariance}
\label{sec:varlike_adap}

\subsection{Adaptive covariance matrix}
\label{sec:varlike_ada}
\begin{figure*}
\sidecaption
\includegraphics[width=12cm]{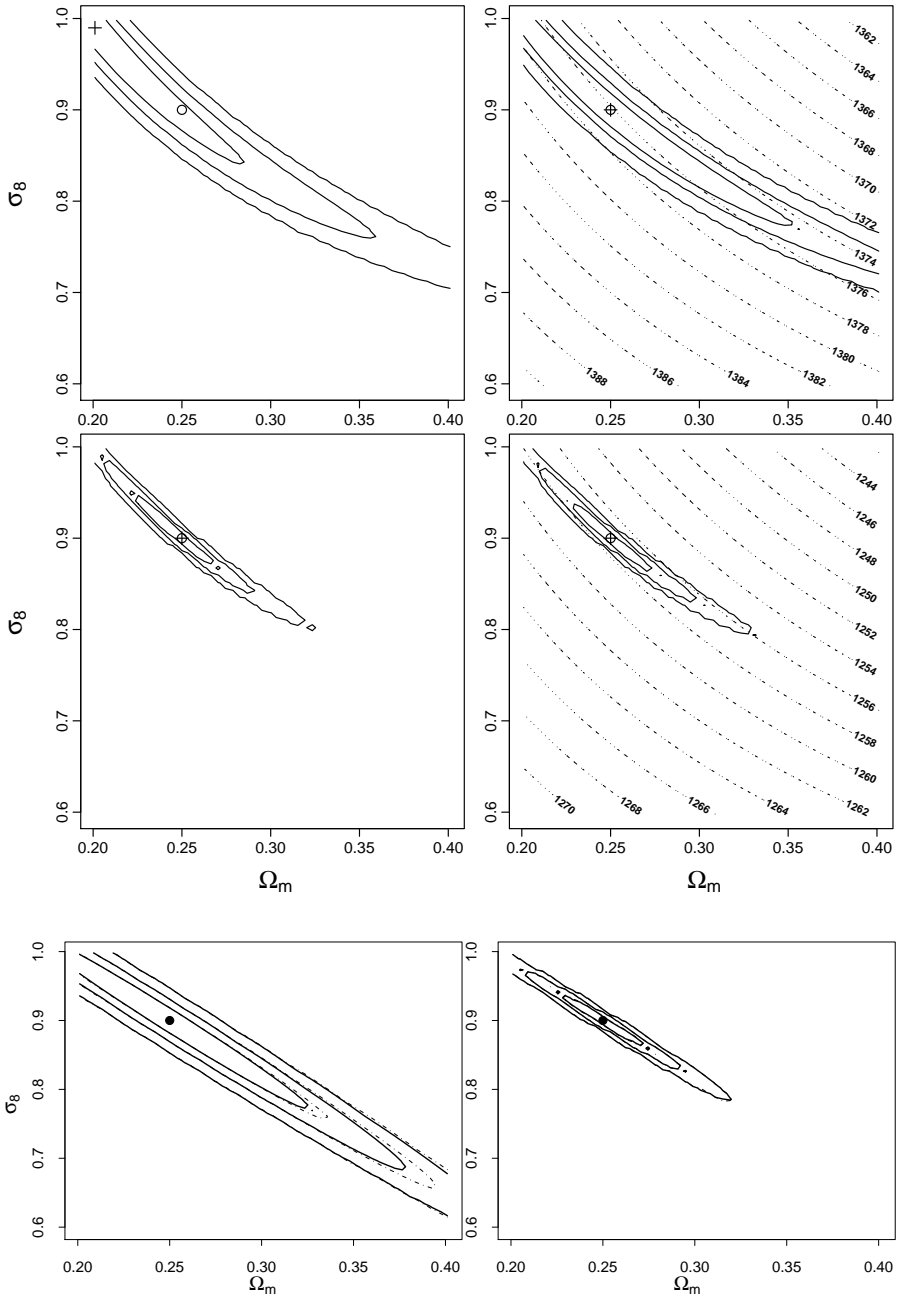}
\caption{The left plots shows the likelihood contours when using a model- dependent covariance, more explicitly, when calculating the posterior from (\ref{eq:likefunc_adap}). The cross illustrates the best-fit value, whereas the circle indicates our fiducial model. The panels on the r.h.s. show the likelihood contours obtained when neglecting the determinant-terms (\ref{eq:likefunc_adap}). The dotted contours visualize regions of constant $\ln |\C_{\pi'}|^{-1/2}$. The likelihood contours in the upper row correspond to a survey size of 84 deg$^2$, whereas the lower panels correspond to $A=900$ deg$^2$.}
\label{fi:like_var}
\end{figure*}

For a given cosmological model we can calculate the covariance directly from (\ref{eq:cov++}) - (\ref{eq:covM}). This enables us to perform a likelihood analysis, where the covariance is calculated individually for every point in parameter space. We denote this parameter dependent covariance as $\C_{\vpi}$ and rewrite the likelihood (\ref{eq:likefunc}) as
\be
\label{eq:likefunc_adap}
p(\vxi |\vpi) = \frac{\exp \left[- \frac{1}{2}\, \left((\vxi_{\vpi} - \vhxi)^\mr t \;\C_{\vpi}^{-1} \;( \vxi_{\vpi} - \vhxi ) \right) \right]}{(2 \pi)^{d/2} \; |\C_{\vpi} |^{\frac{1}{2}}}  \, .
\ee
Compared to the case of a constant covariance, there are two main differences. First, the covariance in the exponential term of (\ref{eq:likefunc_adap}) changes according to the considered point in parameter space. Second, $ |\C_{\vpi} |^{\frac{1}{2}}$ is now parameter dependent, therefore the determinant no longer cancels with a similar term in the evidence. As a consequence, the posterior likelihood does not only depend on the exponential terms, which basically compare $\vxip$ and $\vhxi$, but it is also affected by the determinants of the covariance matrices, more precisely by their behavior in parameter space. In the following we quantify the impact of the determinant term.\\
The upper left panel in Fig. \ref{fi:like_var} shows the likelihood contours for a $84$ deg$^2$ survey, where the posterior probability is calculated via the new likelihood (\ref{eq:likefunc_adap}). For comparison, the right panel shows the likelihood contours when neglecting the parameter dependence in the determinant terms, hence considering a parameter dependent covariance only in the exponential terms. One clearly sees that the determinant terms shift the likelihood contours and cause a difference between the best-fit value and the fiducial model. In order to explain this shift we overlay the right panels of Fig. \ref{fi:like_var} with the contours of constant $|\C_{\vpi}|^{-1/2}$ (for numerical reasons we plot $\ln |\C_{\vpi}|^{-1/2}$).  We see, that  the covariance determinant is a monotonic function of $\om$ and $\sig$; it decreases with increasing $\om$ or $\sig$. Hence, $|\C_{\vpi}|^{- \frac{1}{2}}$ induces a parameter-dependent weighting, which increases the likelihood at small $\om$ and $\sig$ and vice versa suppresses large $\om$ and $\sig$.\\
In general, the exponential term dominates the likelihood, $|\C_{\vpi}|^{-1/2}$ only has significant impact on parameter regions where the exponential hardly changes. For the highly degenerate case of $\om$ and $\sig$, this applies to curves where $\sig \approx const \times \om^{-0.6}$. Compared to these curves, the contours of constant $|\C_{\vpi}|^{-1/2}$ are slightly rotated, which allows for different values of the latter in regions where the exponential term is constant. As a result, the likelihood contours in the left panel are shifted and stretched towards regions of larger $|\C_{\vpi}|^{-1/2}$ compared to the right panel. Note that for a different parameter combination this bias might not cause such a large shift of the best-fit value.\\
The second row of Fig.  \ref{fi:like_var} shows the same analysis but for a $900$ deg$^2$ survey. Comparing the left and right panel, we see that the likelihood contours are, similar to the $84$ deg$^2$ survey, shifted and stretched towards regions of larger $|\C_{\vpi}|^{-1/2}$. However, the effect is hardly noticeable and the bias of the best-fit value has basically vanished. This can be explained when looking at the expression of the posterior likelihood
\be
\label{eq:likefunc_tot}
p(\vpi |\vxi) = \frac{\exp \left[- \frac{1}{2}\, \left((\vxi_{\vpi} - \vhxi)^\mr t \;\C_{\vpi}^{-1} \;( \vxi_{\vpi} - \vhxi) \right) \right]}{\int \mr d \vpi' |  \C_{\vpi'}^{-1} \, \C_{\vpi} |^{\frac{1}{2}} \exp \left[- \frac{1}{2}\, \left((\vxi_{\vpi'} - \vhxi)^\mr t \;\C_{\vpi'}^{-1} \;( \vxi_{\vpi'} - \vhxi) \right) \right] }  \, .
\ee
Compared to the case of a constant covariance the above expression has an additional factor in the denominator, i.e. $|\C_{\pi} \C_{\pi'}^{-1}|^{1/2}$. Note, that this factor is independent of the survey size $A$, whereas the importance of the exponential term increases with increasing $A$. As a result, the cosmology dependence of the covariance determinant becomes negligible for sufficiently large surveys. 

\subsection{Fisher matrix analysis}
We expect tighter constraints on cosmological parameters if the cosmology dependence of both, mean data vector and covariance matrix, is incorporated into the likelihood analysis, instead of only using the mean data vector \citep{tth97}. The Fisher information matrix can be used to illustrate this effect; its definition reads \citep{ks79,tth97} \\
\be
\label{eq:fisherdef}
\mathbf F_{ij}=\ensav{ \frac{\partial^2 L(\vpi)}{\partial \pi_i \, \partial \pi_j}} =\left( \frac{\partial^2 L(\vpi)}{\partial \pi_i \, \partial \pi_j} \right)_{\vpi = \vpi_\mr{ML}} \,,
\ee
where $ L = -\ln p( \vpi| \vxi)$, $\vpi= (\pi_1,...,\pi_n)$ describes the underlying (cosmological) parameters, and $\vpi_\mr{ML}$ denotes the maximum likelihood parameter vector.\\
In addition, if we Taylor-expand $L(\vpi)$ in parameter space at the fiducial parameters we derive  
\be
\label{eq:Taylor_like}
L(\vpi)= L (\vpi_\mr{fid})+ 0 + \frac{1}{2}(\vpi -\vpi_\mr{fid})^\mr t \, \mathbf T^{-1}(\vpi -\vpi_\mr{fid})+ O(\Delta_\pi^3) \,,
\ee 
with
\be
\label{eq:cov_para}
(\mathbf T^{-1})_{ij} =\left( \frac{\partial^2 L(\vpi)}{\partial \pi_i \, \partial \pi_j} \right)_{\vpi = \vpi_\mr{fid}}\,.
\ee 
The first-order term vanishes since $(\partial L/ \partial \vpi)_{|\vpi_\mr{fid}}$ is zero, hence  (\ref{eq:Taylor_like}) is dominated by second-order terms. In this analysis we only consider $\om$ and $\sig$; $\vpi_\mr{ML}$ corresponds to our fiducial model, i.e. $\vpi_\mr{fid}=\left(\om=0.25,\sig=0.9\right)$. Comparing (\ref{eq:cov_para}) and (\ref{eq:fisherdef}) one sees that the Fisher matrix and the inverse parameter covariance matrix $\mathbf T^{-1}$ are equal. We rewrite (\ref{eq:Taylor_like}) as
\be
\label{eq:Taylor_like2}
L(\vpi)= L (\vpi_\mr{fid})+ \frac{1}{2}(\vpi -\vpi_\mr{fid})^\mr t \, \mathbf F \, (\vpi -\vpi_\mr{fid})+ O(\Delta_\pi^3) \, .
\ee 
\begin{figure*}
\sidecaption
\includegraphics[width=12cm]{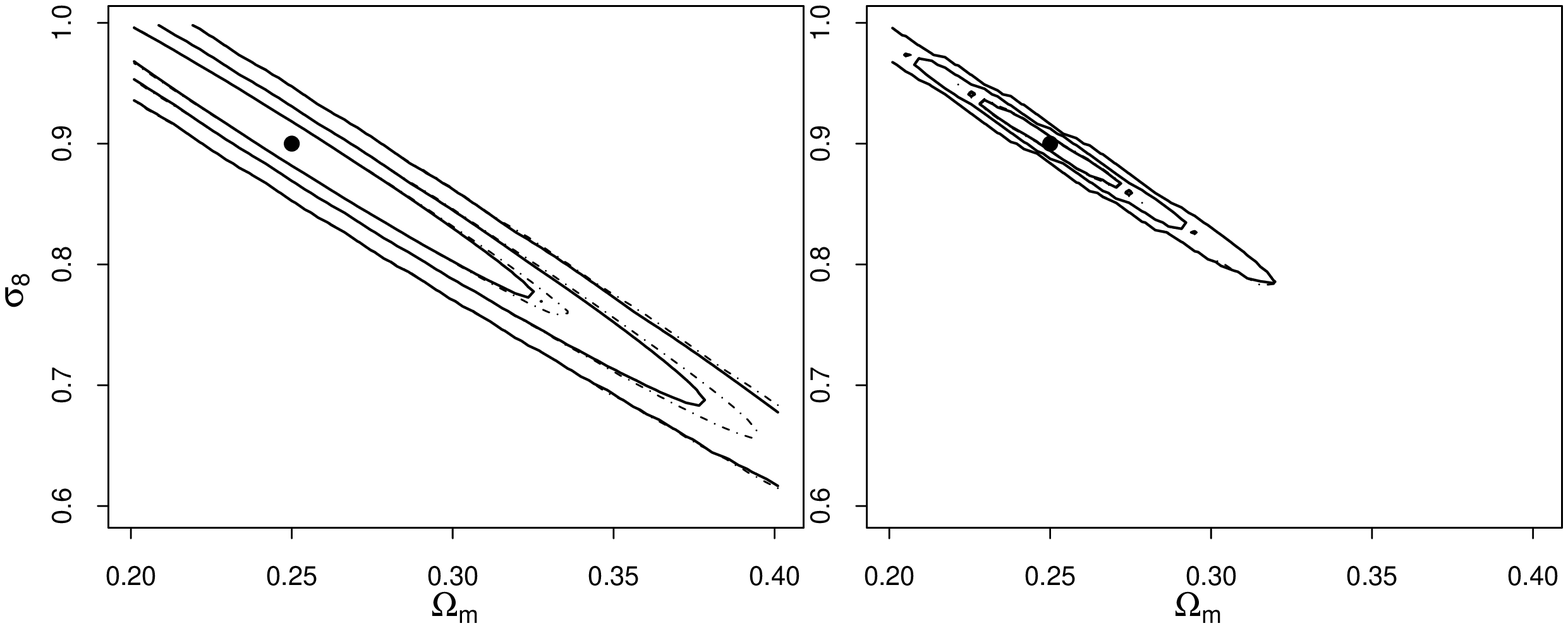}
\caption{Likelihood contours from a Fisher matrix analysis for a 84 $\mr{deg}^2$ survey (\textit{left}), and for a 900 $\mr{deg}^2$ (\textit{right}). The dashed lines correspond to the same analysis but neglecting the covariance term. The dot indicates the fiducial model at which the Fisher matrix was calculated. Note, that in the right panel dashed and solid contours are identical}
\label{fig:fisher}
\end{figure*} 
This equation enables us, for a given Fisher matrix, to calculate lower bounds for $L(\vpi)$, hence we can derive lower bounds on the likelihood contours. For the case that $p(\vxi|\vpi)$ is Gaussian, which at least close to the maximum likelihood parameter vector is a good approximation ($p(\vxi|\vpi) \propto \exp \left[-L\right]$), one can directly express the Fisher matrix in terms of the mean data vector and the data covariance matrix \citep[e.g.][]{tth97}
\be
\label{eq:fishergauss}
\mathbf F_{ij}=\frac{1}{2} \mr{tr} \left[ \C^{-1} \C_{,i} \C^{-1} \C_{, j}+ \C^{-1} \mathbf M_{ij} \right] \, ,
\ee
where $\C_{,i} \equiv \partial \C / \partial \pi_i$ denotes the derivative of the covariance matrix with respect to the $i$-th component of the parameter vector and $\mathbf M_{ij} \equiv \vec \xi_{,i} \vec \xi_{,j}^{\mr t} + \vec \xi_{,j} \vec \xi_{,i}^{\mr t}$. The first term of (\ref{eq:fishergauss}) vanishes in case the covariance matrix is constant in parameter space, the second term vanishes in case of a constant mean. For cosmic shear we have seen that neither the mean data vector, nor the covariance matrix are independent of cosmological parameters, hence when calculating the Fisher matrix both terms must be taken into account. Recall that $\C \propto 1/A$, which also holds for the derivatives $\C_{,i}$, hence the first term is independent of the survey size. The second term increases proportional to the survey volume, therefore the information gain on cosmological parameters, through incorporating the cosmology dependence of covariances, becomes less important for large surveys.\\
Figure \ref{fig:fisher} shows the results of the Fisher matrix analysis for two different survey sizes ($84$ $\mr{deg}^2$ on the left and $900$ $\mr{deg}^2$ on the right). As expected, the left panel (smaller survey) shows a small improvement, which vanishes completely in case of the larger survey (right panel). Nevertheless, one should keep in mind that we only consider Gaussian covariances. The cosmology dependence of the covariance becomes larger for the case of non-Gaussian covariances for the following reason. Non-Gaussianity increases the cosmic variance term, in particular it becomes important on small scales, which are still dominated by shot noise in the pure Gaussian case. As the CDC-effect mainly results from the cosmic variance term, its strength also increases in the non-Gaussian case. A stronger dependence of the covariance on parameters enlarges the first term in (\ref{eq:fishergauss}), which implies that for the case of truly non-Gaussian covariances the improvement on parameter constraints is more significant than shown in Fig. \ref{fig:fisher}. 

\subsection{Iterative likelihood analysis}
\label{sec:varlike_iter}
\begin{figure*}
\sidecaption
\includegraphics[width=12cm]{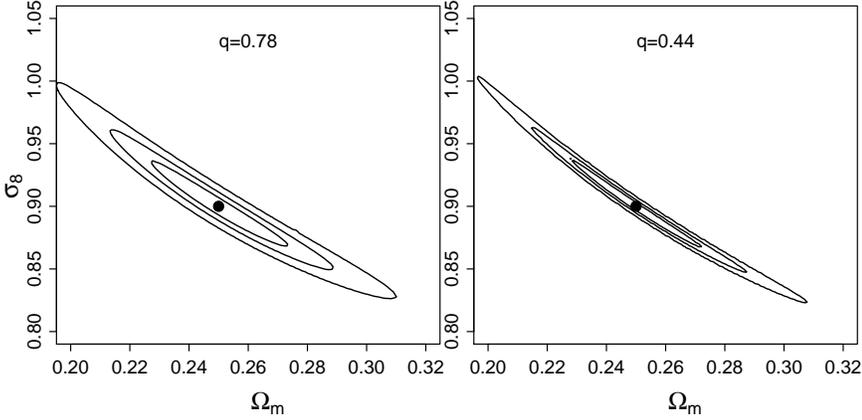}
\caption{The likelihood contours when using a ray-tracing covariance derived from the Millennium Simulation via field-to-field variation (\textit{left panel}), compared to the case of a Gaussian covariance (\textit{right panel}). Although the original size of each field is only 16 $\mr{deg}^2$, we extrapolated the covariance to a 900 $\mr{deg}^2$ survey. The values of $q$ are given in units of $10^{-4}$.}
\label{fig:cov_ray}
\end{figure*}

\begin{table*}
\caption{The ML-parameter sets which occur when choosing different starting cosmologies in the iterative likelihood analysis. }
\centering
\label{tab:iter}
\renewcommand{\arraystretch}{1.1}
\begin{tabular}{c| c c| c c| c c| c c| c c}	\hline \hline
step& \multicolumn{2}{|c|}{run1}& \multicolumn{2}{|c|}{run2} &\multicolumn{2}{|c|}{run3} &\multicolumn{2}{|c|}{run4} &\multicolumn{2}{|c}{run5} \\ 
&$\om$&$\sig$&$\om$&$\sig$&$\om$&$\sig$&$\om$&$\sig$&$\om$&$\sig$ \\ \hline
$\vpi_\mr{start}$ & 0.20 &0.60& 0.237 &0.740& 0.250 &0.90& 0.274 &0.850& 0.40 &1.0 \\
$\vpi_\mr{ML}$ 1& 0.254 &0.892& 0.254 &0.892&0.245  &0.914& 0.259 &0.884& 0.277 &0.858 \\
$\vpi_\mr{ML}$ 2& 0.260 &0.882& 0.260 &0.882& 0.245 &0.914& 0.260 &0.882& 0.259 &0.884 \\
$\vpi_\mr{ML}$ 3& 0.260 &0.882& 0.260 &0.882&  \multicolumn{2}{|c|}{converged}& 0.260 &0.882& 0.260 &0.882 \\
$\vpi_\mr{ML}$ 4&\multicolumn{2}{|c|}{converged}&\multicolumn{2}{|c|}{converged}& \multicolumn{2}{|c|}{converged} &\multicolumn{2}{|c|}{converged}& 0.260 &0.882 \\
\hline
\end{tabular}
\end{table*}
In Sect. \ref{sec:varlike_ada} we have introduced the adaptive covariance, which is a proper way to incorporate cosmology dependent covariances into a likelihood analysis. Its disadvantage is the large computational effort, which is high already for Gaussian covariances. In order to account for non-Gaussianity, one must employ ray-tracing covariances derived from many numerical simulations with different underlying cosmologies. In a multi-dimensional parameter space, this is clearly unfeasible with today's computer power.  \\
In this section we quantify the impact on likelihood contours when using non-Gaussian instead of Gaussian covariances. We use a ray-tracing covariance taken from the Millennium simulation \citep{hhw08}, neglect the CDC-effect and approximate the covariance to be constant in parameter space. The error in the posterior likelihood caused by this approximation increases with increasing distance to the cosmology of the ray-tracing simulation. As we are mainly interested in regions around the maximum likelihood parameter set, $\vpi_\mr{ML}$, this suggests the following strategy for a likelihood analysis. First, perform an iterative likelihood analysis using Gaussian covariances in order to derive $\vpi_\mr{ML}$. Then, start a numerical simulation with this cosmology, derive a ray-tracing covariance, and perform the final likelihood analysis. This ansatz minimizes the errors due to the CDC-effect in the region of interest and additionally incorporates non-Gaussianity. \\
In order to derive $\vpi_\mr{ML}$ iteratively, we start from an arbitrary cosmology, calculate a Gaussian covariance matrix therefrom using (\ref{eq:cov++}) - (\ref{eq:covM}), and perform a likelihood analysis. Throughout this first iteration step the covariance matrix is kept constant. In the second step we choose the ML-parameter set of the first analysis as the underlying cosmology for the new covariance matrix, and again perform a likelihood analysis. We continue this iteration process until the ML-parameter set converges. \\
The main difficulty of this ansatz is that the choice of the starting cosmology might influence the final ML-parameter estimate and therefore also the final covariance. In order to check for this, we take the noise of a ray-tracing data vector, add it to our fiducial data vector and thereby simulate measurement uncertainties in the latter. When performing the analysis without noise the iteration converges after one step, as the model data vector (at the fiducial model) exactly fits the fiducial data vector, $\vec \xi_{\vpi_\mr{fid}}=\vhxi$. Table \ref{tab:iter} shows the results for 5 iterative likelihood analyses, each starting from a different cosmology in the covariance. We see that all 5 runs converge quickly, 4 of them to the same cosmology. Only the run which started from the fiducial model deviates from the others. Although the suggested $\vpi_\mr{ML}$ are close to $\vpi_\mr{fid}$, we note that none of the runs converges to the fiducial model. This implies that the starting cosmology can bias the final outcome of the iterative likelihood analysis and can shift the ML-estimate. In general, such a bias occurs if the function $\vxip - \vhxi$ does not fall off steeply enough around the ML-parameter set, which especially applies to higher-dimensional likelihood analyses. \\
Our iterative pre-analysis has converged to $\om= 0.26$, $\sig=0.882$, however we ``only'' have a ray-tracing simulation with $\om= 0.25$, $\sig=0.9$ available. Figure \ref{fig:cov_ray} shows the result of our likelihood analysis, when using the ray-tracing covariance of the Millennium simulation (left panel).  Compared to a likelihood analysis using a Gaussian covariance (right panel), the contours broaden significantly; $q$ increases from 0.44$\times10^{-4}$ in the Gaussian to 0.78$\times10^{-4}$ in the non-Gaussian case. Note that the value of $q$ in the Gaussian case does not correspond to that in Table \ref{tab:qvalues}, because we use different survey parameters (here, $\sigma_\epsilon=0.3$, $\bar n=15/\mr{arcmin}^2$) and a different data vector (here, 30 logarithmic bins from 0.2-130 arcmin) in order to exactly match the corresponding parameters of the ray-tracing covariance.\\
The impact of non-Gaussianity depends on the scales probed by the data vector. In our case 20 bins are below 10 arcmin, therefore the impact is relatively high. Choosing linear bins or probing higher $\vartheta$ reduces the difference to the Gaussian case. For the data vector considered here, this difference is of the same order as the impact of the CDC-effect we described in Sect. \ref{sec:varlike_para}. However, the strength of the latter will most likely increase for non-Gaussian covariances, as we explained at the end of the last section.
\section{Conclusions}
\label{sec:conc}
An accurate likelihood analysis plays an essential role in future precision cosmology. We can only exploit the full potential of upcoming high quality data, if we use appropriate statistical methods. In this context the derivation of covariances is an important issue in order not to bias the parameter constraints.\\
In cosmic shear, there are several methods to derive covariances. First, one can calculate $\C$ analytically assuming a Gaussian shear field. This assumption breaks down on small angular scales ($< 10$ arcmin), where non-linearities of the matter density field start to become important. Second, covariances can be estimated from ray-tracing simulations. Although computationally more expensive, this method automatically accounts for the non-Gaussianity of the shear field. In both methods the covariance is calculated assuming a specific cosmology. In the first case, this cosmology enters in the power spectrum from which $\C$ is calculated, in the second case we estimate $\C$ from numerical simulations, which are also based on a given cosmology. Past cosmic shear data analyses approximate the covariance to be constant in parameter space, therefore assume that its underlying cosmology does not influence the result of a likelihood analysis significantly.    \\
In this paper we have shown that the covariance matrix depends non-negligibly on its underlying cosmology and that this CDC-effect significantly influences the likelihood contours of parameter constraints. To prove this, we calculate 2500 Gaussian covariance matrices for various parameters of $\om \in [0.2;0.4]$ and $\sig \in [0.6;1.0]$; all other cosmological parameter are held fixed. Even a change of $\om$ and $\sig$ within the WMAP5 68\% confidence levels has a non-negligible impact on the likelihood contours. Here, the value of $q$ deviates by a factor of 1.84 and this deviation increases to 2.76 if one considers the WMAP5 95\% confidence levels. Furthermore, we show that the impact of the CDC-effect depends on survey parameters. Although the likelihood contours become smaller, relatively the CDC-effect becomes more important when increasing the survey size or when decreasing the ratio $\sigma_\epsilon^2/\bar n$.  Therefore, a proper treatment becomes more important in the future, for large and deep surveys.\\
To take cosmology dependent covariances into account we present two methods. First, we perform a likelihood analysis with an adaptive covariance matrix. Here, $\C$ is calculated individually for every point in parameter space, assuming the corresponding parameters as the underlying cosmology. For small surveys this method introduces a bias to the best-fit parameter set, which vanishes when going to larger survey sizes. A disadvantage of this approach is its computational costs. Using the analytic expression for Gaussian covariances is already time-consuming, using ray-tracing covariances to include the non-Gaussianity is not feasible with today's computer power. For the Gaussian case we present a fast and convenient scaling relation to derive covariances on a parameter grid. As a side-effect this approach enhances the constraints on cosmology, for the reason that we now incorporate two cosmology dependent quantities into the likelihood analysis instead of only the mean data vector.  \\
In a strict sense the second method does not account properly for the CDC-effect, however it minimizes the error around the maximum likelihood parameter set ($\vec \pi_\mr{ML}$). The method consists of two steps, first derive $\vec \pi_\mr{ML}$ through an iterative process, then derive a ray-tracing covariance with $\vec \pi_\mr{ML}$ as underlying cosmology and incorporate this in the final likelihood analysis. Here, the approximation of a constant covariance is made, however the error in the posterior probability is minimized in the region of interest; in addition, this ansatz incorporates non-Gaussianity which is non-negligible for future surveys. A drawback is the fact the the starting point of the iteration might bias $\vec \pi_\mr{ML}$. This must be checked carefully before employing this method, otherwise the approximation of a constant covariance fails around  $\vec \pi_\mr{ML}$.


\bibliographystyle{aa}


\begin{acknowledgements}
We thank Ismael Tereno and Martin Kilbinger for useful discussions and advice. This work was supported by the Deutsche Forschungsgemeinschaft under the projects SCHN 342/6--1 and SCHN 342/9--1. TE is supported by the International Max-Planck Research School of Astronomy and Astrophysics at the University Bonn. 
\end{acknowledgements}

\end{document}